\RequirePackage{amsmath}
\documentclass[runningheads]{llncs}
\usepackage[T1]{fontenc}
\usepackage{graphicx}
\usepackage{booktabs}
\usepackage[misc]{ifsym}
\newcommand{\corr}{(\Letter)}

\usepackage{algorithm}
\usepackage{algorithmicx}
\usepackage[noend]{algpseudocode}
\usepackage{mathtools}
\usepackage{subcaption}
\usepackage{xcolor}
\usepackage{amsmath}
\usepackage{amssymb}
\usepackage{cite}
\usepackage{wrapfig}
\usepackage{multirow}
\usepackage{amsthm}

\usepackage{etoolbox}
\newtoggle{camera-ready-version}
\toggletrue{camera-ready-version}

\newtheorem{assumption}[theorem]{Assumption}
\newtheorem*{theorem*}{Theorem}

\begin{document}
\title{Bandits for Sponsored Search Auctions under Unknown Valuation Model: Case Study in E-Commerce Advertising}

\titlerunning{Bandits for SSA: Case Study in E-Commerce Advertising}

%
\iftoggle{camera-ready-version}{
\author{
    Danil Provodin\inst{1} \corr \orcidID{0009-0000-8986-6510} \and
    J\'er\'emie Joudioux\inst{2}\orcidID{0000-0003-0655-9262} \and
    Eduard Duryev\inst{2}}
\authorrunning{D. Provodin et al.}
%
\institute{Eindhoven University of Technology, Eindhoven, The Netherlands
\email{d.provodin@tue.nl}\\ \and
Zalando SE, Berlin, Germany\\
\email{\{jeremie.joudioux, eduard.duryev\}@zalando.de}}
}{
\author{Anonymous Author(s)
\institute{Affiliation}
}
}
\maketitle              
\begin{abstract}

This paper presents a bidding system for sponsored search auctions under an unknown valuation model. This formulation assumes that the bidder's value is unknown, evolving arbitrarily, and observed only upon winning an auction. Unlike previous studies, we do not impose any assumptions on the nature of feedback and consider the problem of bidding in sponsored search auctions in its full generality. Our system is based on a bandit framework that is resilient to the black-box auction structure and delayed and batched feedback. To validate our proposed solution, we conducted a case study at 
\iftoggle{camera-ready-version}{
Zalando, a leading fashion e-commerce company.}
{a leading e-commerce company X.}
We outline the development process and describe the promising outcomes of our bandits-based approach to increase profitability in sponsored search auctions. We discuss in detail the technical challenges that were overcome during the implementation, shedding light on the mechanisms that led to increased profitability.

\keywords{Sponsored search auctions \and Adversarial bandits \and Unknown valuation model \and Batched and delayed feedback.}
\end{abstract}
\section{Introduction}\label{sec1}

Search engine advertising is one of the central topics of the modern literature on e-commerce advertising. Typically, advertising opportunities are sold in search engines through sponsored search auctions (SSA), run separately for each search query and offering ad slots as the entity being sold. Advertisers participating in these auctions repeatedly bid for slots on a search engine, aiming to maximize their profitability from the ads they show. 

SSAs present a unique interplay of challenges, including black-box auction mechanisms, batched and delayed feedback, click attribution, etc. These complexities are further compounded by the common challenge that advertisers often lack a clear assessment of the valuation for the slots being sold at the auctions \cite{feng_2017}. Therefore, learning in SSA becomes exceptionally demanding, requiring solutions that can adeptly navigate through a spectrum of real-life challenges. However, prevailing solutions often resort to simplifications and restrictive assumptions, limiting their effectiveness in practical applications.

This work aims to address these challenges by proposing a comprehensive solution for bidding in SSAs within an industrial context. We consider the bidding problem in its full generality, i.e., we do not impose any assumptions on the nature of feedback or underlying auction mechanisms and consider the value of a good at sale unknown. 

In contrast, a standard assumption in the literature on auction theory is that participants arriving in the market have a clear assessment of their valuation for the goods at sale (see, e.g., \cite{risk_av_agent18, Han_2020a, Han_2020b}). However, this assumption is severely violated in online advertising, where the high-frequency auction mechanism is subject to changing market conditions and user behavior. Additionally, numerous studies assume a structured underlying auction mechanism, with some positing a second-price auction lying at the heart of the allocation process (see, e.g., \cite{weed2015, regmin_spauctions15, NIPS2017_bidding_with_side_inform, RTB2017, alibaba_DQN2018}), while others suggest a first-price auction (see, e.g., \cite{Han_2020a, Han_2020b, han_2022}). In reality, the underlying auction mechanism often remains a black box, as exchanges do not disclose the type of auction mechanism being used. In fact, there has been a surge in the popularity of deep learning-based auction designs in recent years. These designs involve learning an approximation of an allocation function that satisfies specific properties \cite{deep_gsp_2021, neural_auction_2021}.

Further, a list of challenges faced by practitioners in present-day advertising scenarios arises from the interactive nature of SSA. While clicks can be observed shortly after an ad is displayed, it may take hours or even days for the corresponding sale to occur. This \textit{delay in feedback} can make it challenging to optimize bids or other decisions in online advertising, as the true value of action may take time to be apparent \cite{Delays}. Furthermore, since conversion rates are typically low in online advertisement, the \textit{signal provided to the learner is infrequent}, and observational data might not capture behavioral patterns. This scarcity of information makes it difficult for bidders to make informed decisions, complicating the optimization process. Finally, a large volume of multiple-item auctions takes place per iteration, and bidders typically submit a batch of bids. Consequently, bidders have access only to information aggregated over a certain period, leading to \textit{batched feedback} - where the rewards for a group of actions are revealed together and observed at the end of the batch. 
While, these challenges are well-studied in the online learning literature, to the best of our knowledge, we are not aware of any studies that address them from an auction perspective.

To address these challenges, we employ online learning in adversarial bandits \cite{auer_abandits}. The adversarial setting is advantageous for this problem as it does not rely on any structural assumptions regarding the reward components, which is perfectly combined with the black-box nature of SSAs. Next, we introduce the BatchEXP3 algorithm, an extension of the EXP3 algorithm, which is capable of batch update and prone to delayed feedback, and develop a simple yet effective and practical bid placement system.

To validate the effectiveness of the proposed approach, we deployed our bidding system and conducted a three-month live test. The evaluation of our system shows a notable increase in profit during the test duration, suggesting that our approach offers a practical solution to addressing common real-life challenges.

In summary, the paper makes the following contributions:
\begin{itemize}  
    \item We formulate a learning-to-bid problem under an unknown valuation model as a bandit learning task, illustrating how real-life challenges can be adeptly formalized (Section \ref{sec:prob_form}).
    \item We introduce the Batch EXP3 algorithm -- an extension of the EXP3 algorithm designed to handle batched feedback (Section \ref{sec:algorithm}).
    \item We develop an effective bid placement system capable of learning in the black-box auction and handling batch and delayed feedback (Section \ref{subsec:deployment}).
    \item We apply our methodology to a real-world SSA problem, showcasing empirical results that not only highlight the effectiveness of our system but also reveal valuable insights for bidding system designers (Section \ref{sec:experiments}).
    \item We reflect on the practical difficulties we encountered and elaborate on gaps that prevent a seamless transition from theory to practice. (Section \ref{sec:reflection}).
\end{itemize}

\section{Related work}
\label{sec:related_work}
\paragraph{Auctions}
The majority of auction theory research has focused on designing truthful auction mechanisms to maximize the seller's revenue by optimizing a \textit{reserve price} (a price below which no transaction occurs) \cite{optimal_auctions81, optimal_auctions81_2, comp_info69}. A more traditional approach takes a game-theoretic view, where the seller has perfect or partial knowledge of bidders' private valuations modeled as probability distributions \cite{Wilson1985GameTheoreticAO}. This approach is limited as it relies on perfect knowledge of the bidders' value distribution, which is unlikely to be known in practice \cite{risk_av_agent18}. 

The ubiquitous collection of data has presented new opportunities where previously unknown quantities, such as the bidders’ value distributions, can potentially be learned from past observations  \cite{weed2015}. In recent years, data-driven efforts have emerged that aim at learning auction mechanisms to optimize the seller's or bidder's profit \cite{regmin_spauctions15, 10.5555/2887007.2887118, NIPS2014_66368270}, or multiple performance metrics simultaneously, including social welfare and revenue \cite{neural_auction_2021, deep_gsp_2021}. In particular, this line of research has triggered the development of various reinforcement learning (RL) and bandit approaches, which we describe below in more detail.

\paragraph{RL and bandits for bid optimization.} 
The existing research to optimize bidding strategy falls into two main categories: methods based on the bandits framework and methods based on the full RL framework.

The origin of the full RL methods in application to auctions can be traced back to \cite{RTB2017}. In this paper, the authors modeled a budget-constrained bidding problem as a finite Markov Decision Process (MDP). 
The authors utilize a model-based RL setting for the optimal action selection, assuming the perfect knowledge of the MDP. This work led to various improvements, such as proposing model-free RL algorithms \cite{alibaba_DQN2018}, where the learner cannot obtain perfect information, considering continuous action space using policy gradient methods \cite{a2c_RTB}, and modeling the mutual influence between bidders using multi-agent RL framework \cite{Jin_2018, 10.1145/3488560.3498373}. We refer to \cite{bidding_survey} for a comprehensive survey of related works.

One practical downside of RL approaches is that they are known to be sample-inefficient. Moreover, their high complexity renders them challenging to deploy in real-life scenarios due to their limited debuggability. Conversely, bandit-based methods have demonstrated effectiveness in optimizing bidding strategies across various contexts. A substantial body of literature on optimal bidding strategies focuses on efficient budget optimization utilizing bandits and online optimization techniques \cite{10.1145/3442381.3450074, balseiro2023analysis, gao2022bidding}. Particularly, in \cite{10.1145/3442381.3450074}, the focus is on the problem of online advertising without knowing the value of showing an ad to the users on the platform. However, it is noteworthy that the existing works heavily depend on the assumption of a known pricing mechanism and do not generalize to black-box auction mechanisms.

Numerous bandit-based methods have been proposed without budget constraints (see, e.g.,  \cite{weed2015, regmin_spauctions15, NIPS2017_bidding_with_side_inform} for the second-price auctions, \cite{Han_2020a, Han_2020b, han_2022} for the first-price auctions, and \cite{feng_2017} for generalized auction mechanisms). Notably, due to the truthfulness of second-price auctions, methods developed for such a mechanism are based on optimism in the face of uncertainty principle \cite{weed2015, regmin_spauctions15, NIPS2017_bidding_with_side_inform}, whereas first-price and generalized auction mechanisms leverage the adversarial nature of the problem \footnote{First-price and generalized auctions are known to be untruthful, making the environment from the bidders' perspective adversarial.} and use exponential weighting methods \cite{Han_2020a, Han_2020b, feng_2017, han_2022}.

One paper we found particularly relevant to our approach is  \cite{feng_2017}. In this paper, a general auction mechanism is considered, where the product valuation $v_t$ is unknown, evolving in an arbitrary manner and observed only if the bidder wins the auction. The authors decompose the reward of placing bid $b_t$ at iteration $t$ as $r_t(b_t) = rev_t(b_t)x_t(b_t)$, where $x_t(\cdot)$ is the allocation function and $rev_t(\cdot)$ is the revenue function. Consequently, they assume that while $rev_t(\cdot)$ is subject to bandit feedback, $x_t(\cdot)$ is subject to online feedback, i.e., the learner gets to observe $x_t(\cdot)$ for each bid $b$ and not only for the placed bid $b_t$. Based on this, they develop an exponentially faster algorithm (in action space) than a generic bandit algorithm. As we will discuss in the next section, our setting is more complex, and a similar assumption does not apply to us.

\section{Problem formulation}
\label{sec:prob_form}

We focus on a single bidder in a large population of bidders. At the beginning of each round $\tau$, the bidder has a value $v^{\prime}_{\tau} \in \mathbb{R}$ per unit of a good and, based on the past observations, submits a bid $b_{\tau} \in B$, where $B$ is a finite set of bids (will be specified later). The outcome of the auction is as follows: if $x^{\prime}_{\tau}(b_{\tau}) = 1$ (click occurred), the bidder gets a good and pays $p^{\prime}_{\tau}(b_{\tau})$; if $x^{\prime}_{\tau}(b_{\tau}) = 0$ (no click occurred) the bidder does not get a good and pays nothing.

In reality, the sponsored search engine runs multiple auction rounds per iteration and reveals only aggregated information over a set of auctions. Formally, every iteration $t$, $t=1,\dots, T$, where $T$ is unknown and possibly infinite horizon, is associated with a set of reward contests $I_t$. The bidder picks a bid $b_t$, which is used at all reward contests. \footnote{Note, changing a bid within the reward contest $I_t$ would not make any sense, as the bidder does not have access to granular information about every single auction.} At the end of iteration $t$, the bidder observes \textit{aggregated values} of revenue $\sum_{\tau \in I_t} v^{\prime}_{\tau} x^{\prime}_{\tau}(b_t)$, payment  $\sum_{\tau \in I_t} p^{\prime}_{\tau}(b_t) x^{\prime}_{\tau}(b_t)$, and \textit{click-through-rate} $\frac{1}{\lvert I_t \rvert} \sum_{\tau \in I_t} x^{\prime}_{\tau}(b_t)$ in the reward contest $I_t$.

We denote $v_t = \sum_{\tau \in I_t} v^{\prime}_{\tau} x^{\prime}_{\tau}(b_t)$, $p_t(b_t) = \sum_{\tau \in I_t} p^{\prime}_{\tau}(b_t) x^{\prime}_{\tau}(b_t)$ and define the \textit{instantaneous profit} as follows:

\begin{equation*}
    \label{eq:reward_batch}
    \textstyle r(b_t, v_t; p_t) = \sum_{\tau \in I_t} \left ( v^{\prime}_{\tau} - p^{\prime}_{\tau}(b_t) \right ) x^{\prime}_{\tau}(b_t) = v_t - p_t(b_t),
\end{equation*}
and the bidder's goal is to maximize the \textit{total profit} $\max_{b_t \in B} \sum_{t=1}^{T} r(b_t, v_t; p_t)$.

\begin{remark}
$\lvert I_t \rvert$ is a random variable whose distribution is unknown to the learner. Moreover, different allocation and payment functions might be used for different auctions, i.e., $x^{\prime}_\tau(\cdot)$ and $p^{\prime}_\tau(\cdot)$ depend on $\tau$ as opposed to \cite{feng_2017}.
\end{remark}

Next, bidders are rarely presented with a single item, and in real life, they strive to optimize bids for multiple items simultaneously. Let $\mathcal{C}$ be a finite set of possible contexts. Every iteration $t$ is associated with a unique set of contexts $C_t \in \mathcal{C}$. Given $C_t$, the bidder selects a vector of bids $\textit{\textbf{b}}_t \in B^{	\lvert C_t \rvert}$, one for each context, and observes vectors of $\textit{\textbf{v}}_t$ and $p_t(\textit{\textbf{b}}_t)$, where $p_t$ is vector functions from $\mathbb{R}^{	\lvert C_t \rvert}$ to $\mathbb{R}^{	\lvert C_t \rvert}$. The instantaneous profit, in this case, is defined as the inner product between the vector of profits $ \textit{\textbf{v}}_{t} - p_{t}(\textit{\textbf{b}}_{t})$ and vector of ones $\textbf{1}$:


\begin{equation}
\label{eq:reward_final}
    \textstyle r(\textit{\textbf{b}}_t, \textit{\textbf{v}}_t; p_t) = \left \langle \textit{\textbf{v}}_{t} - p_{t}(\textit{\textbf{b}}_{t}),  \textbf{1} \right \rangle.
\end{equation}
and the goal becomes to maximize the \textit{total profit} for multiple items

\begin{equation}
\label{eq:goal2}
    \max_{\textit{\textbf{b}}_t \in B^{\lvert C_t \rvert}} \sum_{t=1}^{T} r(\textit{\textbf{b}}_t, \textit{\textbf{v}}_t; p_t).
\end{equation}

Finally, the bidder does not observe the outcome of bid $b_t$ immediately after reward contest $I_t$ ends. Instead, the outcomes are batched in groups and observed after some delay. 
Formally, let $\mathcal{T} = t_1,...,t_{M}$ be a grid of integers such that  $1 < t_1 < ... < t_{M} = T$, where $M$ is number of batches. It defines a partition $\mathcal{S} = \{ S_1,...,S_{M} \}$ where $S_1 = [1:t_1]$ and $S_k = (t_{k-1}, t_k]$ for $k \in [2:M]$. The set $S_k$ is the $k$-th batch. Next, for each $t \in [T]$, let $J(t) \in M$ be the index of the current batch $S_{J(t)}$. Then, for each $t \in S_{J(t)}$, the bidder observes the outcome of reward contest $I_t$ only after batch $S_{J(t) + \Delta}$ ends, for some positive integer $\Delta$.

Although the bidder's goal \eqref{eq:goal2} remains unchanged in the batched feedback setting, we emphasize that the complexity of the problem increases greatly, as the decision at round $t$ can only depend on observations from $\Delta$ batches ago. In fact, \cite{provodin_batchbandits} shows that, in the worst case, the performance of the batch learning deteriorates linearly in the batch size for stochastic linear bandits.

The instantaneous profitability \eqref{eq:reward_final} and the goal \eqref{eq:goal2} correspond to the most general setting of learning in sponsored search auction when the bidder aims to optimize bids for multiple items simultaneously under the batched feedback. This is the problem that we are addressing in this paper.

\begin{assumption}[Independence of goods at sale]
\label{assumption1}
A set of possible contexts is represented by $n$ unit vectors, $C_t = \{e_1, ..., e_n \}$ for every $t=1,...,T$.
\end{assumption}
Such context set definition corresponds to the situation when the bidder is presented with $n$ items to bid for and treats these items independently. 
Alternatively, \cite{Han_2020a} and \cite{han_2022} propose to use valuation $v_t$ or its estimate as a context. However, we consider a stricter setting, assuming that $v_t$ is unknown for the bidder, nor data is available for its estimation. 

\begin{assumption}[Fixed batch size and delay]
\label{assumption2}
Grid $\mathcal{T}$ divides the horizon $T$ in equal partitions, i.e., $\lvert S_k \rvert = q \quad \forall S_k \in \mathcal{S}$, for some $q > 0$, and delay $\Delta$ is fixed for all rounds. Moreover, values of $q$ and $\Delta$ are known to the bidder in advance.
\end{assumption} 

Although restrictive from the problem formulation perspective, the batch size and delay are controlled by our bidding system and can be wholly justified in practice (see Section \ref{subsec:deployment}). Moreover, our algorithm, which we introduce in Section \ref{sec:algorithm}, is adaptable to unknown and random batch sizes and delays.

\subsection{Learning in sponsored search auctions}
\label{subsec:bandit_form}


Due to the unique characteristics of the problem described above, we take an online learning perspective and formalize the goal in Eq. \eqref{eq:goal2} employing the adversarial bandits setting. Specifically, we assume that the bidder (learner) is presented with a discrete set of bids (actions) $B$. At each auction (round) $t$, the learner picks an action $b_t$ and the adversary constructs reward $r_t$ by secretly choosing reward components $v^{\prime}_{\tau}$, payment functions $p^{\prime}_{\tau}(\cdot)$, and allocation functions $x^{\prime}_{\tau}(\cdot)$, which is further observed by the learner. \footnote{In the subsequent sections, we will use pairs bidder-learner, bid-action, and outcome-reward interchangeably, depending on the context.}

The advantage of the adversarial setting 
is that it avoids imposing any assumptions on the reward components (except that $r_t \in [0,1]$), which is perfectly combined with the black-box nature of sponsored search auctions. Moreover, as we mentioned in Section \ref{sec:related_work}, the adversarial setting allows accounting for the untruthfulness of the underlying auction mechanism. 

The goal of the learner, in this case, becomes minimizing \textit{regret}, the difference between the learner's total reward and reward obtained by any fixed vector of bids in hindsight, which is equivalent to maximizing the total profit in Eq. \eqref{eq:goal2}:
\begin{equation*}
    \textstyle R(T, n) = \sup_{\textit{\textbf{b}} \in B^{n}} \mathbb{E} \left [ \sum_{t=1}^T \left ( r_t(\textit{\textbf{b}}, \textit{\textbf{v}}_t; p_t) - r_t(\textit{\textbf{b}}_t, \textit{\textbf{v}}_t; p_t) \right ) \right ].
\end{equation*}

\section{\textsc{BatchEXP3}: Algorithm for learning in SSA}
\label{sec:algorithm}

We introduce an adaptation of the EXP3 algorithm to the batch setting, which we call Batch EXP3. Batch EXP3 enables strong theoretical guarantees by building on top of the basic algorithm and extensions to more complex settings with batched and delayed feedback. Specifically, Batch EXP3 maintains $n$ instances of the learner, each instance for a separate item, and performs $\Delta$-steps delayed update at the end of each batch. Importantly, the update mechanism of Batch EXP3 preserves the importance-weighted unbiased estimator of rewards, thus, making our adaptation resilient to batch and delayed feedback. Algorithm \ref{algo} formalizes the description above.

Theoretical guarantees of the Batch EXP3 follow from a classical analysis of the EXP3 algorithm (see, e.g., \cite{auer_abandits}). For completeness and theoretical rigor, we formulate a separate statement on Batch EXP3 theoretical performance. 

\noindent
\begin{minipage}{0.47\textwidth}
\vspace{3pt}
    \begin{theorem*}
    \label{thm1:regret_bound}
        The regret of BatchEXP3 algorithm with learning rate $\sqrt{\frac{\log \lvert B \rvert}{T \lvert B \rvert}}$ is: $\mathcal{O} \left ( n \sqrt{ q T \lvert B \rvert \log \lvert B \rvert} + n \Delta \right )$.
    \end{theorem*}
    
    \begin{proof}
    We start analysis with $n=1$:
        \begin{align*}
            & R(T, 1)  =  \sup_{b \in B} \mathbb{E} \left [  \sum_{t=1}^T \left ( r_t(b) - r_t(b_t) \right ) \right ] \\
            & \leq \sup_{b \in B} \mathbb{E} \left [  \sum_{k=1}^M \sum_{t \in S_k} \left ( r_t(b) - r_t(b_t) \right ) \right ] + \Delta \\
            & \stackrel{(a)}{\leq} q \sup_{b \in B} \mathbb{E} \left [  \sum_{k=1}^M \left ( r_t(b) - r(b_t) \right ) \right ] + \Delta \\
            & \stackrel{(b)}{\leq} 2 q \sqrt{M \lvert B \rvert \log \lvert B \rvert} + \Delta \\
            & = 2 \sqrt{q T  \lvert B \rvert \log \lvert B \rvert} + \Delta,        
        \end{align*}
        where $(a)$ is because of Assumption \ref{assumption2}, $(b)$ follows from standard analysis of EXP3, and $r_t(\cdot) = r(\cdot, v_t; p_t)$.
        Then, summing $R(T, 1)$ over $n$ items gives $\mathcal{O} \left ( n \sqrt{ q T \lvert B \rvert \log \lvert B \rvert} + n \Delta \right )$.
    \end{proof}
    \hspace{1cm}
\end{minipage}%
\begin{minipage}{.49\textwidth}
\vspace{-35pt}
  \begin{algorithm}[H]
    \centering
    \caption{\textsc{BatchEXP3} algorithm}
    \label{algo}
    \begin{algorithmic}[1]
    \Require bid set $B$, learning rate $\eta$, number of items $n$, grid $\mathcal{T}$, delay $\Delta$
    \State Set $X^j_{0,i} = 0$ for all $i \in [B]$ and $j \in [n]$
    \State Set $t \gets 1$ 
    \For{$t = 1, 2,  \dots $}
            \State \textbf{\textit{(Policy update)} } 
            \State Compute sampling distribution $\pi_t=(\pi^j_t)_{j=1}^n$: 
            \begin{equation*}
                \pi^j_{t}(b_i) = \frac{\exp{(\eta X^j_{t-1,i})}}{\sum_{l \in [B]} \exp{(\eta X^j_{t-1,l})}}
            \end{equation*}
            \State \textbf{\textit{(Bid generation)} } 
            \State Sample $\textit{\textbf{b}}_t \sim \pi_t(\cdot)$
            \State $X^j_{t,i} \gets X^j_{t-1,i}$
            \If{$t \in \mathcal{T}$ and $J(t) > \Delta$}
                    \For{$s \in S_{J(t) - \Delta}$}
                        \State Observe $\textit{\textbf{v}}_s, p_s(\textit{\textbf{b}}_s)$
                        \State Calculate $r(\textit{\textbf{b}}_s)$ 
                        \State Calculate $X^j_{t,i}$:
                        \begin{equation}
                            X^j_{t,i} = X^j_{t,i} + 1 - \frac{\mathbb{I} \{ b^j_s=b_i \} (1 - r^j(\textit{\textbf{b}}_s)) }{\pi^j_t(b_i)}
                            \label{eq:update_rule}
                        \end{equation}
                        
                    \EndFor
            \EndIf
            \State $t \gets t+1$
    \EndFor
    \end{algorithmic}
    \end{algorithm}
\end{minipage}

\section{Deployment}
\label{subsec:deployment}
While our methodology accounts for the black box auction mechanism, unknown valuation of the goods at the sale, and batch update, it takes system support to fully address click attribution and the data collection issues caused by delayed feedback. For example, Algorithm \ref{algo} simply assumes $\textit{\textbf{v}}, p(\textit{\textbf{b}})$ data is correctly provided as input, but this is nontrivial in practice. To provide a systematic solution, we fill these gaps on the deployment side. Subsequently, we describe the live test design for our bidding system.

\subsection{Bidding system architecture}

\paragraph{Clicks attribution}
It is rarely the case when a single click leads to a desired outcome. Usually, a customer journey starts with a single click, but it is a chain of clicks that results in a conversion event. Identifying click chains and attributing credit to a single click in each click chain is a complex independent task that requires great engineering efforts. 
\iftoggle{camera-ready-version}{
At Zalando, the \textit{performance measurement pipeline} takes up these challenges and measures the performance of online marketing at scale. In short, the pipeline sources all marketing clicks, sales, as well as more complex conversion events such as customer acquisitions, and creates the customers’ journeys across all their devices, from first ad interaction to conversion. Then, it determines how much incremental value was created by every ad click based on a large number of parallel randomized experiments. We refer the interested reader to \cite{zalando_attribution} for more details.
}{
At company X, the \textit{performance measurement pipeline} takes up these challenges and measures the performance of online marketing at scale. In short, the pipeline sources all marketing clicks, sales, as well as more complex conversion events such as customer acquisitions, and creates the customers’ journeys across all their devices, from first ad interaction to conversion. Then, it determines how much incremental value was created by every ad click based on a large number of parallel randomized experiments.
}

\paragraph{Bidding system} 
\begin{wrapfigure}{r}{0.4991\textwidth} 
  \vspace{-31pt}
  \begin{center}
    \includegraphics[width=0.499\textwidth]{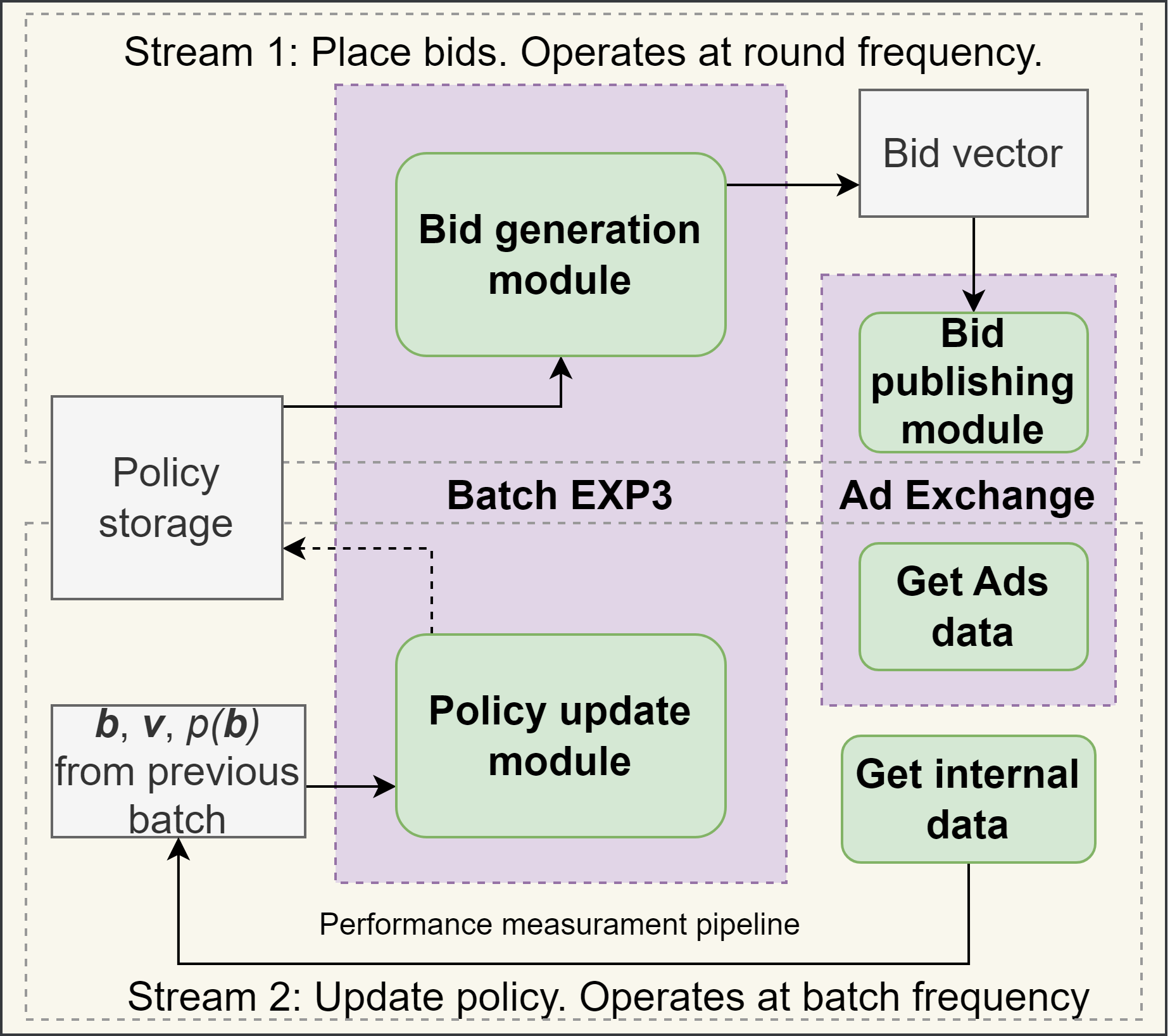}
  \end{center}
  \caption{Bidding system architecture}
  \label{fig:arch}
  \vspace{-13pt}
\end{wrapfigure}

Our bidding system is designed to match the modularity of the bandit methodology in an efficient way, including the \textit{bid generation} and \textit{policy update} components and the \textit{performance measurement pipeline}. The bidding system contains two streams \cite{JSSv094i09}: the first stream is responsible for bid placement in the ad exchange; the second stream unifies the \textit{performance measurement pipeline} and the \textit{policy update} component.  

Ideally, both streams are to be synchronized and run as frequently as possible, one right after the other. 
However, the \textit{performance measurement pipeline} is subject to daily execution due to its compoundness and complexity, making any attempt to increase the frequency of the second stream meaningless. 
Keeping the same frequency for the first stream would admit placing one bid a day, which slows down the learning process considerably. 

To account for this limitation, we desynchronize two streams and execute the first stream with a higher frequency, updating bids every 3 hours. Such improvement allowed us to speed up the learning process substantially. Therefore, the first stream runs every 3 hours and samples bids from the latest policy (3-hour time period corresponds to round $t$), while the second stream is subject to daily execution and performs an update based on batched feedback (scheduled by grid $\mathcal{T}$). \footnote{The first stream runs at 12am, 3am, 6am, etc. The second stream runs at 12am.} The architecture is illustrated in Figure \ref{fig:arch}.

\subsection{Live test design and unfolding}
\label{subsec:test_design}

\paragraph{Test scope} A subset of 180 ($n=180$) clothing products was selected to be steered by the bandits algorithm. Because of the data sparsity, we chose to focus for this test on products for which the traffic was deemed high enough. The selection criterion was that they should meet a threshold of ten average daily clicks over a period of six months. The products selected for the test were randomly sampled amongst those satisfying this traffic threshold.   

\paragraph{Profit metric}
\begin{wrapfigure}{r}{0.43\textwidth} 
  \vspace{-31pt}
  \begin{center}
    \includegraphics[width=0.43\textwidth]{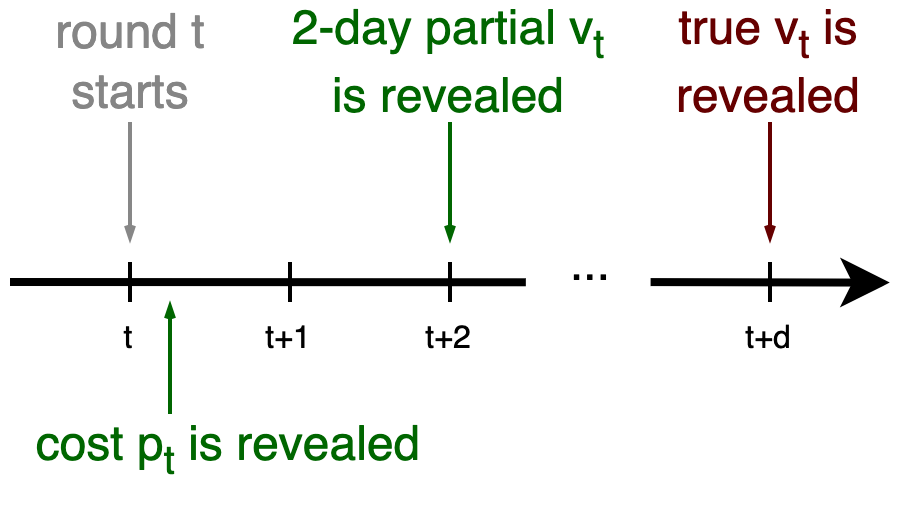}
  \end{center}
  \caption{A timeline of a single round.}
  \label{fig:timeline}
  \vspace{-13pt}
\end{wrapfigure}
In Section \ref{sec:prob_form}, we defined the reward as the aggregated difference between valuation $v_t$ and costs $p_t$. While costs $p_t$ cause no problems and correspond to the expenditure during the round $t$ (which is (almost) immediately available to the bidder), the valuation $v_t$ is abstract and requires special attention. 

We consider valuation $v_t$ unknown to the bidder before auction $t$ starts. In fact, it is difficult to evaluate $v_t$ even when auction $t$ is over. Typically, the ground truth of valuation $v_t$ is assumed to be the revenue auction $t$ has generated over $d$ days, where $d$ might correspond to several months due to return and cancellation policies (see Fig. \ref{fig:timeline} for illustration). Therefore, learning a bidding system when $d$ is too big is impractical. 

Although a vast literature on bandits with delayed feedback provides solutions with delay-corrected estimations, these solutions are not infallible and cannot completely eliminate delay. Therefore, two practical ways of dealing with delays are shortening the feedback loop by decreasing $d$ or developing a delay-free method by substituting $v_t$ with some approximation. Due to the lack of historical data, we have taken a more pragmatic approach and shortened the delay to 2 days. As a result, we focus on maximizing the \textit{2 days partial profit}, i.e., profit attributed within the 2-day conversion window after the bid placement. 

\paragraph{Profit metric normalization}
We apply two normalization steps to the \textit{2 days partial profit}. The first normalization step is a naive yet pragmatic way of incorporating side information into the modeling, and it eliminates the difference between time periods. Since users' activity is different during the nighttime and daytime, rewards that we observe from 3am - 6am are incomparable to rewards that we observe from 3pm - 6pm. We reduce rewards to a common medium of expression by normalization:

\begin{equation*}
    r(b_t, v_t; p_t) \coloneqq \alpha_{t-qS_{J(t)}} r(b_t, v_t; p_t), 
\end{equation*}
where $t-qS_{J(t)}$ is the time period number within the batch $S_{J(t)}$, and $\alpha_l$ is the ratio of average traffic during time period $l$ to the average traffic of the most active time period, $l=1, \dots, q$. 

Next, the bandit formulation requires rewards to be in the $[0,1]$ range, which is far from the truth in a real-life application. In order to account for that, we apply minimax normalization to rewards by

\begin{equation*}
     r(b_t, v_t; p_t) \coloneqq \frac{r(b_t, v_t; p_t) - r_{min}}{r_{max} - r_{min}},
\end{equation*}
where $r_{min}$ and $r_{max}$ are the 5-th and 95-th quantities of the historical \textit{2 days partial profit}. The coefficients $\alpha_l$ were calculated on the market level and remain constant for all products $i=1, \dots, n$; whereas the coefficients $r_{min}$ and $r_{max}$ are product-dependent and were calculated individually for each product based on the limited historical data available.

\paragraph{Bid space}
The bandit formulation described in Section \ref{subsec:bandit_form} supports a discrete and finite set of bids. 
The conducted analysis of historical data demonstrated that bids higher than 40 cents are unprofitable, which narrows the potential bid space values to range from 1 cent to 40 cents. To trade off the total number of bids and coverage of the bid space, we decided to include more options for lower bids (with step 2 cents) and fewer options for higher bids (with steps 3-5 cents). The final bid space $B$ consists of 14 possible bids and is
\begin{equation}\label{eq:bidspacedef}
B = \left\{1,\,3,\, 5,\, 7,\, 9,\, 11,\, 13,\, 15,\, 17,\, 20,\, 25,\, 30,\, 35,\, 40\right\}.
\end{equation}

\paragraph{Test reset}
We started the experiment with a generic value of the learning rate $\eta = 1$. After we rolled out the solution, we spotted unstable learning behavior in the bidding system. Two weeks later, we decided to reset the test with the learning rate $\eta = 0.1$, which corresponds to less aggressive exploitation by the learner. While this adjustment did not resolve the issue completely, it mitigated the level of instability and facilitated the learning process. We will detail this phenomenon in the next section.

\section{Experimental results and discussion}
\label{sec:experiments}

\begin{figure}[tb]
     \centering
      \begin{subfigure}[b]{0.495\textwidth}
         \centering
         \includegraphics[width=\textwidth]{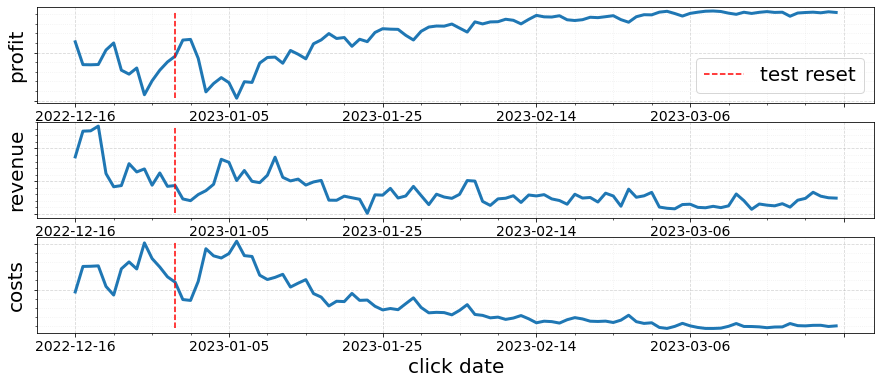}
         \caption{All products.}
         \label{fig:genera_behavior}
     \end{subfigure}
     \hfill
     \begin{subfigure}[b]{0.495\textwidth}
         \centering
         \includegraphics[width=\textwidth]{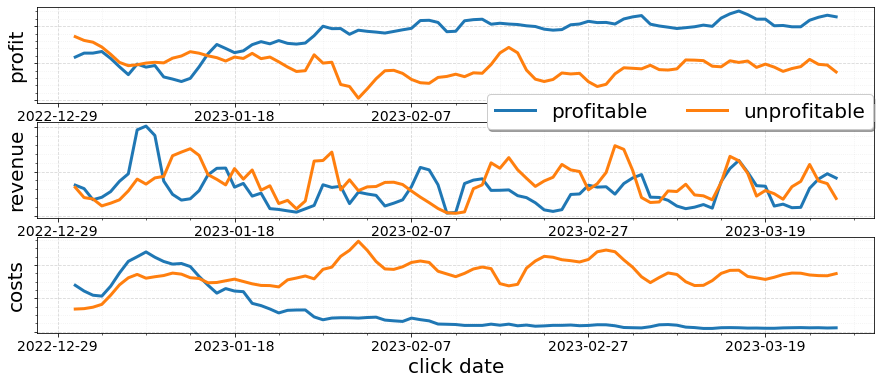}
         \caption{Profitable vs. Unprofitable products.}
         \label{fig:good_bad_skus}
     \end{subfigure}
        \caption{Dynamics of the profit (top), revenue (middle), and costs (bottom).}
\end{figure}

The test took place from December 16th, 2022 (date of deployment) to March 25th, 2023, in a large European country. Figure \eqref{fig:genera_behavior} illustrates the aggregated dynamics of partial profit (top), revenue (middle), and costs (bottom) for a subset of 180 products. To protect business-sensitive information, we have concealed the $y$-axis and only report the relative behavior. As we can see, beginning in January, the system displays a robust positive trend in profit. Notably, there are two peaks in costs that conclude in mid-January, corresponding to exploration periods within our system.  Following January 25, the revenue maintains relative stability, while costs continue to decrease, resulting in an upswing in profitability.

These observations suggest that the increased profitability can be attributed to two mechanisms. Firstly, there is a significant reduction in costs during the initial part of January (until January 25) and the ensuing decline in revenue, which marks the conclusion of the (aggressive) exploration phase. The second mechanism involves a further reduction in costs while sustaining a relatively stable revenue, which suggests bids steering toward more profitable values.

To better understand when the increased profitability is due to selecting bids to increase revenue and when it is the case due to decreased costs, we provide a closer look at business metric behavior on a group level. 

\subsection{Group level analysis}

For identifying product groups, we focus on profit behavior during which the revenue is relatively stable between February 1 and March 25. Most of the products (100 products) align with the general trend illustrated in Figure \ref{fig:genera_behavior}, placing them into either ``profitable'' or ``emerging'' groups. A subset of 67 products are categorized as ``low-traffic'' due to their minimal associated activity. Finally, there are 13 products that exhibit a counterintuitive behavior, characterized by decreasing profits over time. Table \ref{tab:taxonomy} summarizes statistics on the group level.

\begin{wrapfigure}{r}{0.5\textwidth}
\vspace{-51pt}
    \begin{minipage}{0.5\textwidth}
      \begin{table}[H]
\begin{center}
\begin{minipage}{\linewidth}
\caption{Percentage of clicks, costs, and gain partitioned by product groups}\label{tab:taxonomy}%
\resizebox{\textwidth}{!}{%
\begin{tabular}{@{}lllllll@{}}
\toprule

Product &  &  &  &  & average \\
group & \multirow{-2}{*}{\# products} & \multirow{-2}{*}{\% clicks} & \multirow{-2}{*}{\% costs} & \multirow{-2}{*}{\% revenue} & bid value \\
\midrule
profitable               & 14 (8\%)   &  $4.0\%$   &  $2.5\%$  & $14.7\%$ &  5.8\textcent  \\
emerging                 & 86 (48\%)  &  $73.6\%$  &  69.8\%   & $62.4\%$ &  8.3\textcent \\
low-traffic              & 67 (37\%)  &  $4.1\%$   &  $3.2\%$    & $3.9\%$ &  6.1\textcent \\
unprofitable             & 13 (7\%)   &  $18.4\%$    & $24.6\%$    & $19.1\%$ &  17.9\textcent \\
\bottomrule
\end{tabular}}
\end{minipage}
\end{center}
\end{table}
\vspace{-43pt}
    \end{minipage}
  \end{wrapfigure}

It is essential to address the characteristics of the unprofitable group. Among these products, 7 exhibited an excessively large range of $r_{min}$ and $r_{max}$ coefficients, resulting in reduced system sensitivity to environmental signals. Conversely, for 6 products, $r_{min}$ and $r_{max}$ coefficients were too narrow, causing the system to be overly sensitive to environmental signals. This disparity in coefficient ranges with respect to each product led to a discrepancy between the original signals from the environment and the rewards observed by the agent. Consequently, this misalignment resulted in confounding updates and an inaccurate valuation of bid values. 

\paragraph{Profitable vs Unprofitable product groups} 
Although the unprofitable group primarily results from technical issues within our system, it offers valuable insights when compared to the profitable group. Figure \eqref{fig:good_bad_skus} presents the profit, revenue, and cost dynamics for profitable and unprofitable product groups. 
Notably, the profitable group shows a consistent increase in profit, accompanied by a continual decrease in costs. In contrast, the unprofitable group exhibits a stable or even declining trend in profit, with costs remaining relatively stable.

This evidence suggests that the system optimized bids towards higher profitability due to decreased costs rather than increased revenue. In fact, it was spending almost 10 times less on profitable products than on unprofitable products to accumulate a comparable amount of revenue (14.7\% vs 19.1\%). Nevertheless, it is important to emphasize that, due to the auction nature of the problem, revenue is inherently monotonic to costs, and we cannot expect an increase in revenue with decreasing costs. At the same time, we can see that keeping costs unchanged did not result in a positive outcome for the unprofitable group.

\noindent\fbox{\begin{minipage}{\linewidth}
\paragraph{Takeaway 1} The increasing profitability of our system primarily stems from cost reduction rather than revenue growth.
\end{minipage}}

\paragraph{Low vs high bid values}
To better understand if our system's reduction in costs was a rational reaction, we provide further insights on revenue growth for different bid values, aiming to compare products that converged to lower bid values and those that converged to higher bid values. We focus on profitable and emerging product groups (100 products) and disregard unprofitable (as those are deemed erroneous) and low-traffic (due to unperceptive behavior) groups.

\begin{figure}[tb]
     \centering
     \begin{subfigure}[b]{0.469\textwidth}
         \centering
         \includegraphics[width=\textwidth]{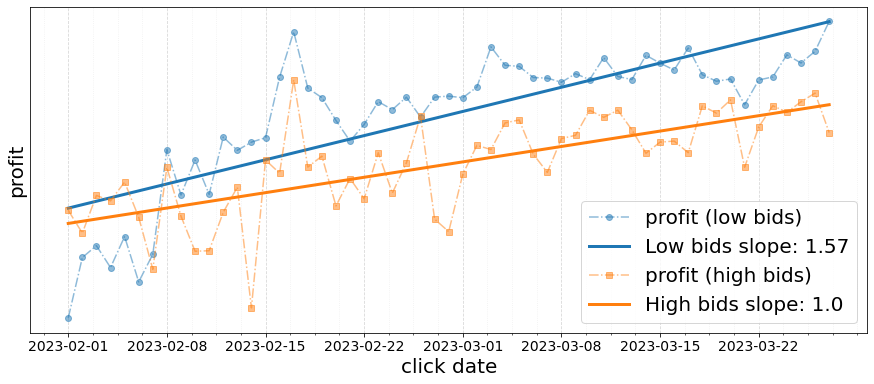}
         \caption{Low vs. high bids.}
         \label{fig:roi_high_low}
     \end{subfigure}
     \hfill
      \begin{subfigure}[b]{0.469\textwidth}
         \centering
         \includegraphics[width=\textwidth]{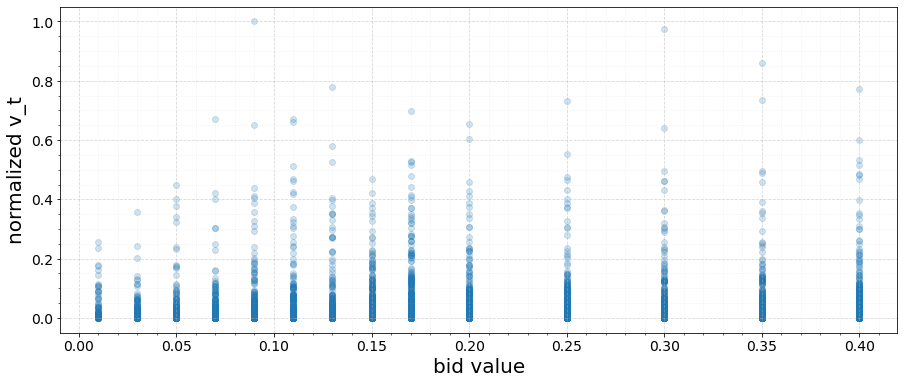}
         \caption{$v_t$ vs bid values.}
         \label{fig:bids_vs_pc2}
     \end{subfigure}
     \hfill
     \begin{subfigure}[b]{0.469\textwidth}
         \centering
         \includegraphics[width=\textwidth]{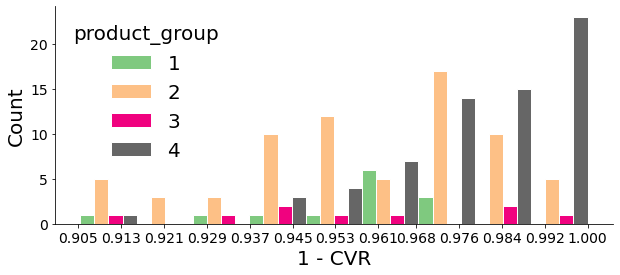}
         \caption{Inverse conversion rate.}
         \label{fig:sparse_ratio}
     \end{subfigure}
        \caption{\eqref{fig:roi_high_low} Profit dynamics for products with higher (than 10 cents) and lower (than 10 cents) bid values, aggregated over highly profitable and profitable product groups. To preserve business-sensitive information, regression slopes represent the normalized linear regression fit. \eqref{fig:bids_vs_pc2} Distribution of normalized $v_t$ across the bid values.. \eqref{fig:sparse_ratio} Distribution of inverse conversion rate categorized by product group: 1 -- profitable, 2 -- emerging, 3 -- unprofitable, 4 -- low-traffic.}
        \label{fig:three graphs}
\end{figure}

Figure \ref{fig:roi_high_low} illustrates the profit dynamics for products with lower bid values (lower than 10 cents) and products with higher bid values (higher than 10 cents). The lower bid values group encounters 75 products, and the higher bid values group encounters 25 products. We can see an upward trend in profit dynamics for both groups, consistent with the observation from Figure \ref{fig:genera_behavior}. Notably, the lower bid value group grows 1.5 times faster than the higher bid value group, as can be seen from regression slopes. Thus, we can conjecture that the system steers bids towards more optimal values by decreasing the costs.

\noindent\fbox{\begin{minipage}{\linewidth}
\paragraph{Takeaway 2} 
The decrease in costs is a rational behavior of our system.
\end{minipage}}

\subsection{Risk of decreasing costs}
While we have seen that reduced costs is a justified behavior, taking a closer look at the low-traffic product group in Table \ref{tab:taxonomy} shows the opposite side of low costs. Specifically, we can see that spending a similar amount of money for a profitable group barely generates any revenue for the low-traffic group. In this part, we discuss the risks associated with the decreasing costs of our system.

Figure \ref{fig:bids_vs_pc2} demonstrates the distribution of normalized $v_t$ across bid values. Specifically, while we observe a slight positive dependence on bid value, values of $v_t$ mostly concentrate around 0 with rare positive signals for all bids. In situations when conversions are rare events, costs become the dominating term in the update rule \eqref{eq:update_rule}, and bids with higher costs are depreciated. This incentivizes the system to spend less rather than earn more, making exploration of higher bid values improbable. As a result, after a certain amount of exploration, the system has ``killed'' 1/3 of products, which could have shrunk marketing opportunities.\footnote{We emphasize that the loss-based update rule \eqref{eq:update_rule} is an inseparable part of the auction problem, and defining a reward-based estimator (instead of loss-based), first, might not make any sense for the auction problem (as trivially bidding the highest value would correspond to the highest revenue), and, second, would lead to similar issues.}

Somewhat surprisingly, this observation goes against one line of research that exploits an assumption of action space structure in the auction problem. This assumption is referred to as the censored feedback structure and says ``Once we know the payoff under a certain bid, then irrespective of whether this bid wins or not, we know the payoff of any larger bid.'' \cite{Han_2020b}. 

\noindent\fbox{\begin{minipage}{\linewidth}
\paragraph{Takeaway 3} Sparse feedback might create asymmetry in the exploration process and make it more intricate to explore bids equally. 
\end{minipage}}

\section{From practice back to theory: Additional insights}
\label{sec:reflection}

Our work presents an adaptation of (theoretically) well-studied results to the industry application. While doing so, we have encountered certain problem-specific difficulties (such as decreasing costs and asymmetric exploration) that are usually overlooked in common theoretical formulations. We believe it is important to understand the roots of these difficulties in order to (1) mitigate them as much as possible with the current theoretical advancements and (2) identify gaps that prevent a seamless transition from theory to practice.

There are three common assumptions in the bandit literature: (i) bounded rewards, often with support in $[0,1]$; (ii) a ``nice'' shape of reward distribution, e.g., sub-Gaussian or Bernoulli; (iii) specific hyperparameters values, optimized in a way to achieve the best theoretical performance.

While these assumptions are not secret, theoretical papers usually focus on ``nicer'' settings and do not explicitly describe the consequences of their violation. However, none of these assumptions are met in the real-world auction problem, and we have encountered difficulties with all of them. Below, we make some observations about what did not work in practice and share some of our learnings.

\paragraph{Bounded rewards} One limitation of many bandit algorithms is that they are typically analyzed assuming rewards fall within the $[0,1]$ range. Our approach employed a direct and practical method to scale the rewards (as outlined in \ref{subsec:test_design}). However, due to the sparsity of rewards and limited data, we encountered challenges in determining the correct $r_{min}$ and $r_{max}$ coefficients for some products. Consequently, this resulted in 7\% of products being unprofitable in our experiment due to the disparity of $r_{min}$, $r_{max}$ coefficients ranges.

Such practical matters are rarely the main focus of theoretical studies \cite{Cowan2017NormalBO}, and no definitive solutions are widely acknowledged. As a practical approach, we recommend adaptive computation of $r_{min}$ and $r_{max}$ for more accurate scaling.

\paragraph{Reward distribution} 
Beyond reward scaling, theoretical performance guarantees usually require the reward to be ``nicely'' distributed. However, this is far from the reality we face in marketing applications. As we can see from Figure \eqref{fig:sparse_ratio}, which reports the inverse conversion rate ($1-CVR$) categorized by groups, the extreme values mostly correspond to low-traffic or unprofitable products, which suggests that striking the right balance between exploration and exploitation becomes even more acute when the inverse conversion rate tends to 1.

In sparse feedback, the EXP3-based update rule \eqref{eq:update_rule} risks assigning very small probabilities to bids with high losses. This is the main reason of decreasing costs, which led to the formation of the low-traffic product group. One way to mitigate this issue is the Exp3-IX-based update rule \cite{NIPS2014_25b2822c, neu2015explore}. However, this would lead to tuning one additional parameter, which, as we discuss later, is highly non-trivial.

\paragraph{Hyperparameters}
The EXP3 family of algorithms requires knowledge of horizon $T$ to get the optimal learning rate value (which is $\eta = \sqrt{log(k) / Tk}$ for classical EXP3 \cite{lattimore_szepesvari_2020}). In cases when $T$ is unknown, theoretical approaches propose either using a doubling trick \cite{besson2018doubling} or decreasing the learning rate. However, the doubling trick is primarily a technical component designed to overcome specific constraints in proofs rather than serving as a practical tool. Decreasing the learning rate, on the other hand, is a more elegant solution \cite{zimmert2022tsallisinf}, but choosing the right schedule for this parameter is not a straightforward task, especially with the sparse reward. 

A slight misspecification of the learning rate can disrupt the entire learning process. At the beginning of the live test, the learning rate was inaccurately set at a generic value of 1, overlooking the sparsity of rewards. This caused heightened sensitivity in the learning system, where omitting to place a bid for 1-2 days resulted in a massive degeneration of policies around this bid value (with probabilities of other bids being compressed towards zero). Consequently, we reset the test and adjusted the learning rate to 0.1. To further address this issue, we would use the decreasing learning rate $\eta_t = 1/\sqrt{t}$ suggested by \cite{zimmert2022tsallisinf}, but we emphasize that it has not been studied from a practical standpoint.

\section{Conclusion}
\label{sec:conclusion}
In this paper, we introduced a systematic solution to learning optimal bidding strategies in black-box sponsored search auctions. Doing so, we have presented how many real-life challenges can be addressed by a synthesis of the bandit framework and our bidding architecture. Live experiment results show the effectiveness of our system in increasing the partial profit, highlighting that the bandit framework is a promising approach to tackle the complex auction problem. Further, our insights highlight the practical difficulties that lurk around this problem, providing practical guidance for bidding system designers.

Naturally, the simplicity of our system has its limitations. We acknowledge that in its current form, our system has no leverage to hinder the diminishing cost effect that primarily arises due to the long-standing problem of sparse reward signals. Exploring this direction further is a promising avenue for future work.

\iftoggle{camera-ready-version}{
\section*{Acknowledgments}
We wish to thank Joshua Hendinata and Aleksandr Borisov for their engineering support.  Furthermore, we would like to thank Amin Jamalzadeh, head of the Traffic Platform Applied Science and Analytics at Zalando, for guidance, support with administrative processes related to the project, and the review of the final draft. Danil Provodin would like to thank Maurits Kaptein and Mykola Pechenizkiy, whose thoughts influenced his ideas.
}{}

\bibliographystyle{splncs04}
\bibliography{main}

\end{document}